\documentclass[aps,prd,letterpaper,11pt,twoside,tightenlines,nofootinbib,showpacs,preprint,twocolumn]{revtex4}
\usepackage{graphicx}
\usepackage[sort&compress]{natbib}
\usepackage{latexsym}
\usepackage{epsfig}
\usepackage{hyperref}

\newcommand{\be}{\begin{equation}}

\newcommand{\ee}{\end{equation}}
\newcommand{\bea}{\begin{eqnarray}}
\newcommand{\eea}{\end{eqnarray}}
\newcommand{\bef}{\begin{figure}}
\newcommand{\eef}{\end{figure}}
\newcommand{\bce}{\begin{center}}
\newcommand{\ece}{\end{center}}
\def\lsim{\mathrel{\rlap{\lower4pt\hbox{\hskip1pt$\sim$}}
    \raise1pt\hbox{$<$}}}         %less than or approx. symbol
\def\gsim{\mathrel{\rlap{\lower4pt\hbox{\hskip1pt$\sim$}}
    \raise1pt\hbox{$>$}}}         %greater than or approx. symbol

\begin{document}

\title{Effects of ultra-light dark matter on the gravitational quantum well}
\author{Paolo Castorina}
\email{paolo.castorina@ct.infn.it}
\affiliation{Dipartimento di Fisica, Universit\`a di Catania,
 and INFN Sezione di Catania, Via Santa Sofia 64
I-95123 Catania, Italia}
\author{Alfredo Iorio}
\email{iorio@ipnp.troja.mff.cuni.cz}
\affiliation{Faculty of Mathematics and Physics, Charles University, V Hole\v{s}ovi\v{c}k\'ach 2, 18000 Prague 8, Czech Republic}
\author{Michal Malinsk\'y}
\email{malinsky@ipnp.mff.cuni.cz}
\affiliation{Faculty of Mathematics and Physics, Charles University, V Hole\v{s}ovi\v{c}k\'ach 2, 18000 Prague 8, Czech Republic}
\date{June 30th 2017}
\begin{abstract}
We study the influence of a periodic perturbation of the effective masses of the nucleons, due to the assumed semi-classical ultra-light dark matter background, on the motion of neutrons in a  gravitational quantum well. Our focus is on the transition probability between the lowest two energy states, with the Rabi frequency in the kHz region corresponding to the series of ``sweet spot'' dark matter masses in the $10^{-11}$eV ballpark. The relevant probability is written in terms of the specific mass and of the effective coupling to the ordinary matter. These parameters can be constrained by the non-observation of any significant deviations of the measured transition probabilities from the dark-matter-free picture.
\end{abstract}
 \maketitle

\section{Introduction}
It is well known that most of the direct detection searches~\cite{Liu:2017drf} for weakly interacting massive particles (WIMPs), perhaps the most popular ~\cite{generalDM} hypothetical form of the Dark matter (DM), may in a not-so-distant future hit their conceptual sensitivity floor corresponding to the irreducible background due to neutrino interactions inside the detectors. Remarkably enough, the room is narrowing down quickly not only for the ``classical'' WIMPs with masses at the level of tens of GeV or above~\cite{Akerib:2016vxi} but, with facilities like SuperCDMS~\cite{Agnese:2016cpb} on the horizon, the same is likely to happen also for much lighter candidates with sub-GeV masses.
In view of that a lot of attention has recently been  paid to alternatives to WIMPs, corresponding to very light or even ultra-light DM (ULDM) candidates such as, e.g.,  axions~\cite{Marsh:2015xka} or other types of scalar fields~\cite{Hui:2016ltb} with masses reaching deep inside the sub-eV region.

However, the interactions of such a substance with the ordinary matter is likely to be very different from the usual particle scattering picture for WIMPs. Indeed, the observed energy density in DM yields occupation numbers so high that the system acts coherently resembling a classical wave rather than a set of individual quanta. Hence, in looking for the effects of such a form of matter in laboratory-based experiments one may take the advantage of simple quantum mechanical systems and consider their response to the quasi-classical DM background.

Recently the sensitivity of the atomic interferometry to oscillating scalar ULDM has been analyzed in \cite{geraci} for scalar fields $\phi$ of masses in the range
$10^{-24}$ eV $\le m_\phi \le 1$ eV, corresponding to the Compton frequency, $f_\phi = m_\phi c^2 / h$, in the range of $10^{-10}$Hz $\le f_\phi \le 10^{14}$Hz. 
Indeed, the linear and quadratic couplings \cite{geraci} of the Standard Model (SM) fields with the DM fields give a modulation of the fermion masses and of the fundamental constants. This affects the mass of particles and of the Earth, the former taken into account by $m_0 \rightarrow m_0+\delta m $, the latter by a modification of the local gravitational acceleration $g_0 \rightarrow g_0+\delta g$.

In this letter we study how a Gravitational Quantum Well (GQW) experiment can either detect or constraint these effects due to ULDM, through a resonance mechanism between the ULDM oscillations frequencies and the frequencies associated to the neutron bouncing phenomenon in the GQW \cite{pignol}. High precison GQW experiments have already been shown to give strong constraints on new physics, such as spacetime noncommutativity \cite{paoloandorfeu}, violation of the equivalence principle~\cite{x1}, and more.
%cosmological constant~\cite{x2},

%The plane of the letter is the following: in Sec. 1 we briefly recall the GQW system; in Sec.2 we discuss the modification due to ULDM couplig with SM fields;
%and the effect on the GQW setting; Sec.3 contains our conclusions.

\section{The Gravitational Quantum Well}

\begin{figure}
\begin{center}
\includegraphics[width=0.5 \textwidth]{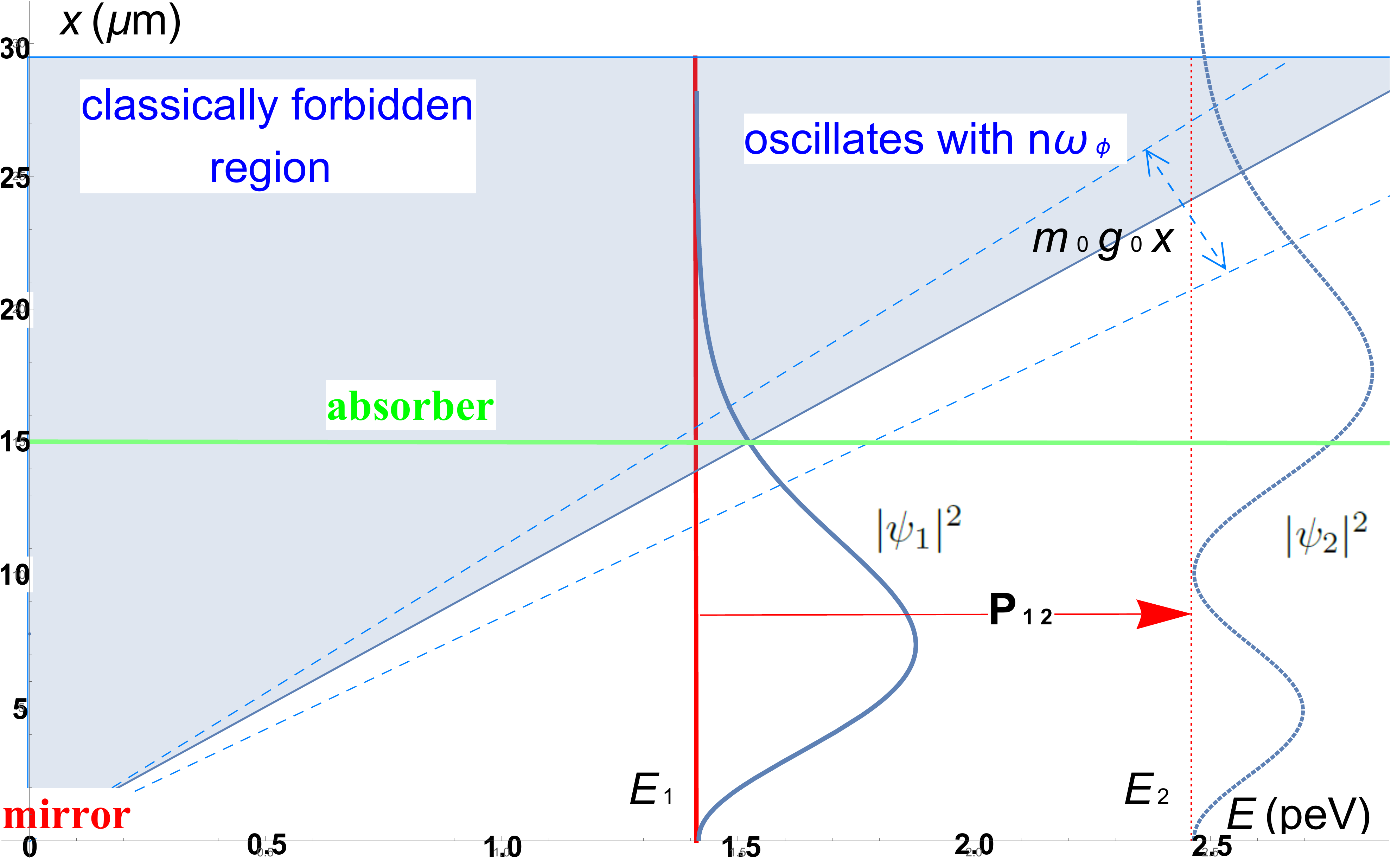}
\end{center}
\caption[3pt]{{\protect\small {A neutron with energy $E_n$ is absorbed when the slit aperture $h$ is equal or less than the corresponding classical turning point $x_s = E_s / m_0 g_0 = - \alpha_s x_0$. In the picture, $x_1 = 13.7 \mu$m, and $x_2 = 24.0 \mu$m, $E_1 = 1.407$ peV, $E_2 = 2.461$ peV. The probability $|\psi_s|^2$ is maximal not at $x_s$, but rather at a smaller value, that in the picture are $x_1^{max} = 7.74\, \mu$m, for $|\psi_1|^2$, and $x_2^{max} = 18.0\, \mu$m, for $|\psi_2|^2$ . Therefore, setting $h$ at about $15\, \mu$m (green line) neutrons in the state $\psi_2$ should not be seen unless a Rabi transition, $\psi_1 \to \psi_2$ induced by ULDM, with probability $P_{12}$,  takes place. The second energy level, $E_2$, and the probability function, $|\psi_2|^2$, are drawn in dots.}}}%
\label{illustration}%
\end{figure}

The GQW is a system conventionally made of a quantum particle in a potential well realized by i) a homogeneous gravitational field with its gradient oriented in the (by definition) vertical direction, say $x$, and ii) a horizontal mirror along, say, $y$ (usually placed at $x=0$), where the particle experiences perfect elastic reflection, see, e.g., \cite{flugge}, and also Fig.~\ref{illustration} here.

The eigenvalue equation, $\hat{H} \Psi_s=E_s\Psi_s$, $s = 1,2,...$, has well known separable form $\Psi_s (x,y) = \psi_s(x) \, \chi(y)$. Here the eigenfunctions corresponding to $x$ are those of bound states (as for any potential well) and, as well known~\cite{flugge}, are given in terms of the Airy function $\varphi$
\be \label{Airy_1}
\psi_s(x) = A_s \varphi(x/x_0 + \alpha_s)~,
\ee
where the $\alpha_s = \{ -2.338, -4.088, -5.521, ...\}$ identify the zeroes of $\varphi$, $x_s = - \alpha_s x_0$, with $x_0 \equiv (\hbar^2 / (2m_0^2g_0))^{1/3}$. One can introduce a dimensionless coordinate $z \equiv x/x_0 + \alpha_s$, in terms of which the normalization coefficients are written as $A_s \equiv (x_0 \int_{\alpha_s}^{+\infty} dz \varphi^2(z))^{-1/2}$. The eigenvalues are
\be \label{Airy_2}
E_s = - m_0 g_0 x_0 \, \alpha_s~.
\ee
For $m_0 \simeq 939.5\,$MeV, the mass of the \textit{neutron}, and $g_0 \simeq 9.81$m/$s^2$, one has $x_0 \simeq 5.87 \mu$m, and $m_0 g_0 x_0 \simeq 0.603$peV.

On the other hand, as the particle is free in the horizontal direction $y$, $\chi (y)$ corresponds to a packet of plane waves of continuous energy spectrum
\be \label{Airy_5}
\chi(y)=\int_{-\infty}^{+\infty}g(k)e^{iky}dk~,
\ee
where $g(k)$ determines the shape of the packet in phase space.

In GQW experiments with neutrons \cite{nesvizhevskyprd2003, Nesvizhevsky:2005ss}, it has been possible to identify their quantum states, $\psi_s(x)$, by realizing a horizontal slit with the upper boundary corresponding to a scatterer/absorber, above the horizontal mirror. When the absorber is at a height less than a critical value, $h < h_s^{crit}$, the neutrons shot into the slit with energy $E_s$ (and greater) do not make it out on the other side of the apparatus, as they are absorbed by the scatterer/absorber. This critical value corresponds to the classical turning point for that given quantized energy\footnote{The extent to which the Equivalence Principle can be said to hold in this experiment is discussed in \cite{pignol}. In this letter we do not consider any such violations, i.e., for us the inertial and gravitational masses are indistinguishable.}, that is, $h_s^{crit} \equiv E_s / m_0 g_0 = - \alpha_s x_0 = x_s$. Detailed description of the experimental set-up can be found in \cite{nesvizhevskyprd2003}, and in the review \cite{pignol} (see also Fig.~\ref{illustration} here), while the report of the first identification of the lowest quantum state is in \cite{nesvizhevskyprd2003}.

Recently, Nesvizhevsky \emph{et al.} \cite{Nesvizhevsky:2005ss} were able to measure the critical heights for the first two quantum states, obtaining the following results
\bea \label{Nesvizhevsky_1}
x_1^{exp}&=&12.2\pm1.8(syst)\pm0.7(stat)\ (\mathrm{\mu m}),\nonumber\\
x_2^{exp}&=&21.6\pm2.2(syst)\pm0.7(stat)\ (\mathrm{\mu m}).
\eea
The corresponding theoretical values can be determined from $x_n = - \alpha_n x_0$ for $\alpha_1=-2.338$ and $\alpha_2=-4.088$, and $x_0 = 5.87 \mu$m yielding $x_1=13.7\:\mathrm{\mu m}$ and $x_2=24.0\:\mathrm{\mu m}$, corresponding to the energy eigenvalues $E_1=1.407\:\mathrm{peV}$ and $E_2=2.461\:\mathrm{peV}$. These values are contained in the error bars, and allow for maximum absolute shifts of the energy levels with respect to the predicted values:
\bea \label{Nesvizhesky_2}
\Delta E_1^{exp}&=&6.55\times10^{-32}\ \mathrm{J}=0.41\ \mathrm{peV},\nonumber\\
\Delta E_2^{exp}&=&8.68\times10^{-32}\ \mathrm{J}=0.54\ \mathrm{peV}.
\eea
In this experiment, neutrons exhibited a mean horizontal velocity of $\langle v_y\rangle\simeq6.5\:\mathrm{ms^{-1}}$.

%\begin{widetext}
%\end{widetext}

\section{ULDM effects}

The interactions of DM fields $\phi$ with the SM matter can be described  by the effective lagrangian densities
\be \label{interaction}
-{L}_{n}^{int}  =  \left( \frac{\sqrt{\hbar c}}{\Lambda _{n,f}}\right )^{n} 
m_{f} \bar{\psi }_{f} \psi _{f}  \phi^{n},
\ee
where $n$ indicates the order of the interaction, $f$ stands for the type of the SM matter under consideration and $m_{f}$ and $\psi _{f}$ denote its masses and
field operators, respectively. All these structures are weighted by the
inverse of the relevant high-energy scales, $\Lambda _{n ,f}$, that also include the a-priori unknown couplings.

The main implication of (\ref{interaction}) is the space- and time-dependent modulation of the fermion masses in the ULDM background
\be
\frac{m_{f}^{eff}}{m_{f}}=  1+\frac{\sqrt{\hbar c} \phi(\vec{r},t)^{n}}{\Lambda_{n,f}^{n}} \,,
\ee
and, in turn, the variation of the local gravitational acceleration (due to the modulation of the mass of the Earth) assuming the DM field permeates through  the Earth body, thus making its mass change slightly in time \cite{geraci}.

To evaluate how this affects the GQW we consider the specific case proposed in \cite{geraci} 
and assume that the local gravitational acceleration and the mass of the neutron both vary periodically in time as
\begin{eqnarray}
g(t)&=& g_0+g_1 \cos{(\omega t}), \label{gt}\\
m(t)&=& m_0+m_1 \cos{(\omega t)} , \label{mt}
\end{eqnarray}
where $\omega=n \omega_\phi$, with $\omega_\phi = m_\phi c^2/ \hbar$.

The motion along the $x$ axis is governed by the Hamiltonian $\hat{H}= \hat{p}^2_x / (2m) + mg \hat{x}$ which, by (\ref{gt}) and (\ref{mt}), becomes
\begin{eqnarray}
\hat{H} & = & \frac{\hat{p}^2_x}{2(m_0+m_1 \cos(\omega t))}  \\
  & + & (m_0+m_1 \cos(\omega t)) (g_0+g_1 \cos(\omega t)) \hat{x} \,. \nonumber
\end{eqnarray}
Since the corrections due to ULDM interaction are small, i.e. $m_1\ll m_0$ amd $g_1\ll g_0$, we can write
$\hat{H} \simeq \hat{H}_0 + \hat{V}_1(t)$, with $\hat{H}_0 \equiv \hat{p}^2_x / 2m_0 + m_0 g_0 \hat{x}$,
$\hat{V}_1(t) \equiv \hat{V}_1  \cos(\omega t)$ and
\be
\hat{V}_1 =  - \frac{\hat{p}^2_x}{2m_0} \frac{m_1}{m_0}
+  \left(\frac{m_1}{m_0} + \frac{g_1}{g_0}\right) m_0 g_0 \hat{x} \,, \label{Vt}
\ee
that are the expressions we shall consider in evaluating the ULDM effects.

\subsection{Time independent corrections}

Just for curiosity let us recall that, for $n=1$, the oscillation frequency in terms of the ULDM mass $m_\phi$ is given by
\be
f_\phi = 2 \pi \omega = 2.4\times 10^{14} \left({m_\phi}{[eV]}\right){\rm Hz}
\ee
and that the time of flight of neutrons in the GQW studied in \cite{pignol} is $T \simeq 40$ ms.
Obviously, for $\omega T \ll 1$, i.e. $m_\phi \ll 6.8 \times 10^{-13}$ eV,  one can neglect the time dependence in $\hat{V}_1(t)$ and the mere effect of the ULDM consists in the shifts in the
energy levels of the unperturbed Hamiltonian $\hat{H}_0$, $E_1$ and $E_2$. For $m_\phi < 10^{-13}$eV these can be readily evaluated by looking at $\langle s|\hat{V}_1|s \rangle$, $s=1,2,...$, with
$\hat{V}_1$ in (\ref{Vt}), and, in order to be compatible with the measurement, they should be within the maximum allowed shifts $\Delta E_1$, $\Delta E_2$ of Eq. (\ref{Nesvizhesky_2}). The numerical calculation, reported in the Appendices, gives
\be  \label{eigenvalue_1}
E_1\left[\frac{m_1}{m_0} 0.341 + \frac{g_1}{g_0} 0.67\right] \le 0.41 \phantom{...} {\rm peV}
\ee
\be  \label{eigenvalue_2}
E_2\left[\frac{m_1}{m_0} 0.333 + \frac{g_1}{g_0} 0.667\right] \le 0.54 \phantom{...}{\rm peV}
\ee
where $E_{1,2}$ are the unperturbed eigenvalues.
Clearly the inequalities (\ref{eigenvalue_1}) and (\ref{eigenvalue_2}) give no strong bound on such ULDM couplings.

\subsection{Time dependent corrections}
For longer transition times or for higher ULDM Compton frequencies (i.e., larger $m_{\phi}$) the time dependence of $\hat H$ can not be neglected. Moreover, the time variation of $\hat V_{1}(t)$ may stimulate  efficient  transitions among different eigenstates leading, eventually, to much stronger limits.
In particular, the $1 \rightarrow 2$ transition probability is governed by the notorious Rabi formula~\cite{Sakurai}
\be
P_{1 2} = \frac{\Omega^2}{\Omega^2 + \delta \omega ^2} \, \sin^2\left(\frac{\sqrt{\Omega^2 + \delta \omega ^2}}{2} \, t\right)
\ee
where $\delta \omega^2 \equiv \omega^2 - \omega^2_{12}$, with $\omega_{1 2} = 2 \pi f_{12}$,
\be
f_{12} \equiv \frac{E_2-E_1}{h } \simeq {\rm 254} \;  {\rm Hz}\,,
\ee
the characteristic frequency, $t$ the time, and
\be \label{OmegaMatrix}
\Omega = \frac{1}{\hbar} \langle 2|\hat{V}_1|1\rangle.
\ee
Combining Eqs. (\ref{Vt}) %and the expression of the eigenfunctions for $n=1,2$ in terms of Airy functions (see Eq.
and (\ref{Airy_1})
%)%
 one gets (see Appendix~\ref{appB})
\be \label{Omega}
\Omega = \frac{g_0 m_0 x_0}{\hbar} \left(\frac{m_1}{m_0}(I_1+I_2)+ \frac{g_1}{g_0} I_1\right)
\ee
where
\bea %\label{I1}
I_1&= &\!\!x_0 A_1 A_2 \int^\infty_{\alpha_1} \!\!dz \varphi(z-\alpha_1+\alpha_2) (z-\alpha_1) \varphi(z),\nonumber\\
\label{I2}
I_2&= &x_0 A_1 A_2 \int^\infty_{\alpha_2} dz \varphi(z-\alpha_2+\alpha_1) z  \varphi(z) \;.
\eea
Since, in the case of our interest, $m_0 g_0 x_0 \simeq 0.603$~peV (see discussion after Eq.(\ref{Airy_2})) one obtains
\be
\Omega[{\rm Hz}] = 914.6 \times  \delta_m
\ee
where
\be
\delta_m =\frac{m_1}{m_0}(I_1+I_2)+ \frac{g_1}{g_0} I_1\,.
\ee

Therefore, one can think of 
%use the GQW device with an absorber placed in a way to
preparing the system so that only the ground state of energy $E_1$ is populated \cite{nesvizhevskyprd2003, Nesvizhevsky:2005ss}. Then, within a given time of flight (which, for  ultra-cold neutrons with $\langle v_y\rangle\simeq6.5\:\mathrm{ms^{-1}}$, is typically of the order of a tenth of a second but, in settings with  reflective vertical mirrors on the edges of the main horizontal one, it may be stretched significantly) there is a finite probability for the system to jump to the first excited state $\psi_{2}$. The corresponding probability $P_{12}$ is, as usual, maximalized at the resonance, $\delta \omega^2 =0$, i.e. at $\omega \doteq 1596$Hz, corresponding, for $n=1$, to $m_\phi \simeq 4\times 10^{-11}$eV (scaling properly for higher $n$) and obeys
\be
P^{\rm max}_{12} = \sin^2\left(\frac{\Omega t}{2}\right) \simeq \frac{\Omega^{2}}{4} \times t^2 \simeq 2 \times 10^5 \phantom{..} \delta^2_m
\ee
for $t = 1$ s and $\delta_{m}^{2}\ll 10^{-5}$. Therefore, an experimental limit on $P_{12}$ gives a bound on $\delta^2_m$.

Concerning the apparent smallness of $P_{12}^{\rm max}$ for more realistic values of $\delta_{m}$ two comments are in order. First, with reference to Fig.\ref{illustration}, one can think of a detector that can distinguish between the $\psi_{1}$ and $\psi_{2}$ states (with some efficiency), placed at the end of the apparatus. Then, although the transition probability $P_{12}$ is small, even a handful of observed events of the $\psi_{2}$ type may constitute the desired signal. Second, after a DM quantum $\phi$ has been absorbed by the neutron, inducing the transition $1 \to 2$, it may then be emitted again, inducing the transition $2 \to 1$ (with  $P_{21} = P_{12}$, see Appendix~\ref{appB} ). On the other hand, the probability of the absorbtion and the subsequent emission is  $P_{12} P_{21} = P_{12}^2$ and, hence, it is strongly suppressed.

\section{Conclusions} The ultra-light dark matter is an intriguing hypothesis that has recently attracted a lot of attention as one of the most interesting alternatives to the notorious WIMP paradigm. Different in many aspects (in particular, high level of coherence in its interactions with matter), its laboratory-based searches may often be performed in table-top experiments focusing on the response of simple quantum-mechanical systems on the presence of the corresponding quasi-classical background perturbations.

In this study, we focus on one of such systems, namely,  the gravitational quantum well which has been recently subject to an intensive laboratory study aimed at measuring the energies of the lowest-laying bound states of neutrons bouncing off the horizontal mirror.
In particular, we focus on the influence of a periodic perturbation modelling the variations of the effective gravitational acceleration due to the assumed semi-classical ULDM background on the transition probabilities of the lowest energy (quasi-)stationary neutron states passing through the apparatus. With the Rabi frequency corresponding to the resonance in the ground to the first excited state transition amplitude in the kHz region (leading to a series of ``sweet spot'' DM masses $m_{\phi}[{\rm eV}]\sim 4\times 10^{-11}/n$ with integer $n$) we rewrite the relevant probability as a function of the DM mass and effective coupling to the ordinary matter. These parameters can be, subsequently, constrained from the the non-observation of any  deviations of the measured transition probabilities from their theoretical ULDM-free spectrum.
\begin{figure}
\begin{center}
\includegraphics[width=0.45 \textwidth]{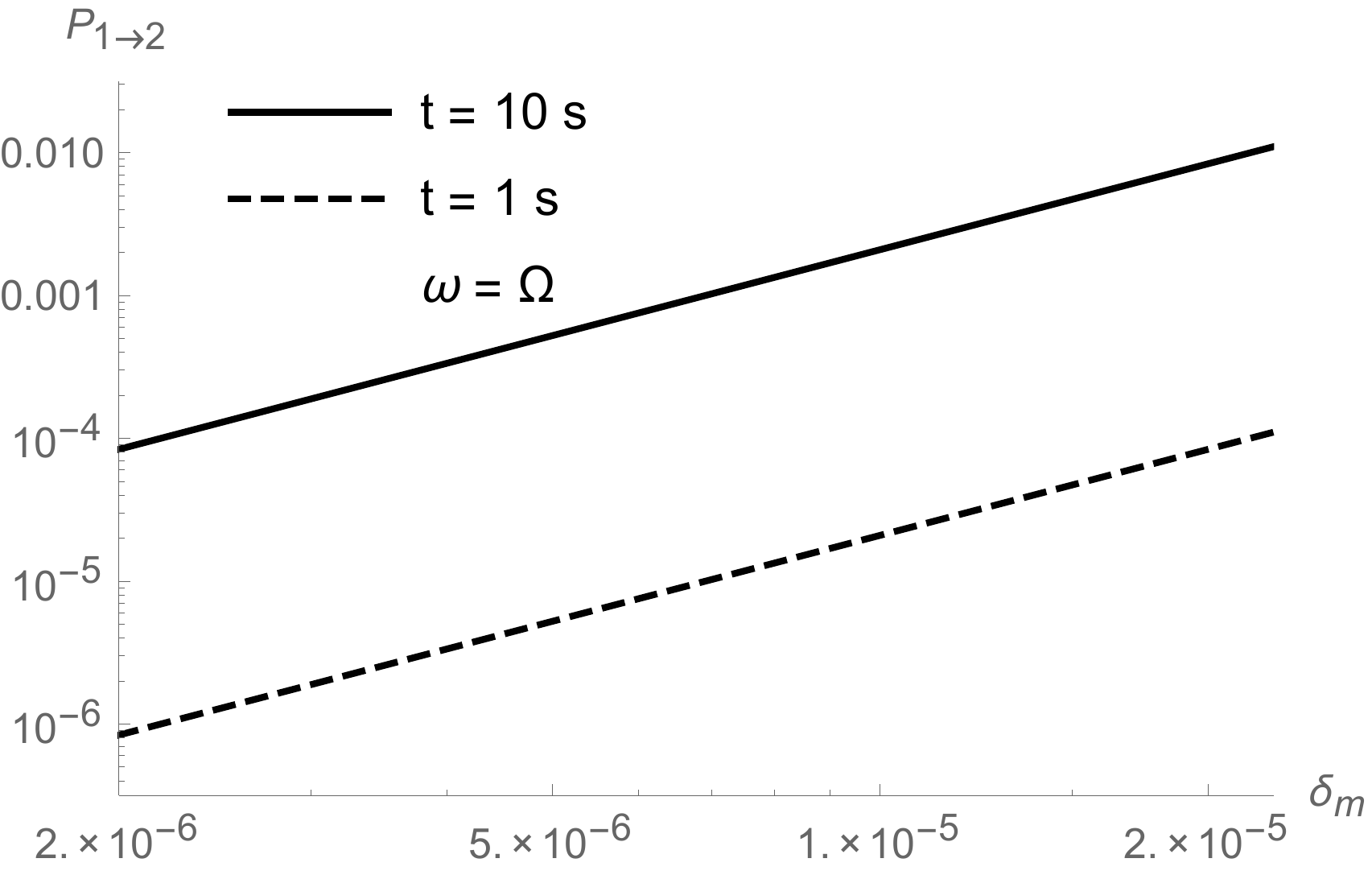}
\end{center}
\caption[3pt]{{\protect\small {Transition probability vs $\delta_m$ at the resonance $\delta \omega=0$, i.e. for $m_\phi= 4.18\times10^{-11}$ eV (provided $n=1$; for higher $n$ the $m_{\phi}$ down-scales accordingly), for two different times of flight, $t = 1$ s and $t=10$ s. For a  specific choice of   $P_{12}$ the allowed region for the effective dark matter coupling $\delta_{m}$ stretches below the corresponding curve.}}}%
\label{first}%
\end{figure}
%
%\begin{figure}
%\begin{center}
%\includegraphics[width=0.4 \textwidth]{MMpicture2.pdf}
%\end{center}
%\caption[3pt]{{\protect\small {***Transition probability vs $\omega$ for $\delta_m = 10^{-5}$ and $t=10$s. UNITS FOR $\omega$ MISSING. ***}}}%
%\label{second}%
%\end{figure}
%
%

\section*{Acknowledgments} A. I. and M.M. acknowledge financial support from the Grant Agency of the Czech Republic (GA\v{C}R), contracts 14-07983S and 17-04902S.

\appendix
\section{Time independent corrections}
The time independent corrections require the calculation of the matrix elements $\langle s|\hat{V}_1|s\rangle$ with $s=1,2,...$ and $\hat{V}_1$ in Eq.(\ref{Vt}), i.e.
\be
\langle s|\hat{V}_1|s\rangle =
\int_{0}^\infty \int_{0}^\infty d x d x' \langle s| x \rangle \langle x| \hat{V}_1 | x' \rangle \langle x'| s \rangle
\ee
where the eigenfunctions are real, $\psi^*_s (x) = \langle s| x \rangle = \langle x| s \rangle = \psi_s(x)$, and given by the Airy function, $\psi_s(x)= A_s \varphi(x/x_0 + \alpha_s)$.
In particular
\begin{eqnarray}
\langle s|\hat{x}|s\rangle & = & \int_{0}^\infty d x \psi_s (x) x \psi_s (x) \nonumber \\
& = & A_s^2 \int_{0}^\infty d x \, x \, \varphi^2 (x/x_0 + \alpha_s)   \nonumber \\
& = & A_s^2 x_0^2 \int_{\alpha_s}^\infty \varphi^2(z) (z-\alpha_s) dz \nonumber \\
& = & x_0 (R_s -\alpha_s) \;,
\end{eqnarray}
where
\be
 R_s= \left(\int_{\alpha_s}^\infty \varphi^2(z)  dz\right)^{-1} \int_{\alpha_s}^\infty \varphi^2(z) z dz
\ee
Since the unperturbed eigenvalues are given by
\be
E_s = - \alpha_s m_0 g_0 x_0 \,,
\ee
the first term of the correction turns out to be
\begin{eqnarray}
\langle s| m_0 g_0 \hat{x} \left(\frac{m_1}{m_0} + \frac{g_1}{g_0}\right)|s\rangle \nonumber \\
= E_s \left(\frac{m_1}{m_0} + \frac{g_1}{g_0}\right) [1 - R_s/\alpha_s]
\end{eqnarray}

Analogously, by using the property of the derivatives of the Airy function, $\varphi''(z) = z \varphi (z)$, the second term of the correction yields
\be
\langle s| -\frac{\hat{p}^2}{2m_0}\frac{m_1}{m_0}|s\rangle = \frac{\hbar^2}{2 m_0}\frac{m_1}{m_0} (m_0 g_0 x_0)^2 R_s
\ee
Putting all together, the final result is
\begin{eqnarray}
\langle s|\hat{V}_1|s\rangle & = & E_s \bigg[\frac{m_1}{m_0}\left( 1- 2 \frac{R_s}{\alpha_s}\right) +\frac{g_1}{g_0}\left (1 - \frac{R_s}{\alpha_s}\right)\bigg]\nonumber\\
\end{eqnarray}
which, by numerical evaluation of $R_s$, gives (\ref{eigenvalue_1}) and (\ref{eigenvalue_2}).

\section{Transition probability\label{appB}}

For the evaluation of the probability of the transition $1 \to 2$, $P_{12}$, one needs the matrix element $\Omega$ in eq.(\ref{OmegaMatrix}).
To compute the correction due to the position dependent term one needs (see before)
\begin{eqnarray}
&& \langle 2| \hat{x}|1\rangle   =  \int_{0}^\infty d x \psi_2 (x) x \psi_1 (x) \label{symmx} \\
& & = A_1 A_2 \int_{0}^\infty d x \, \varphi \left(\frac{x}{x_0} + \alpha_2\right) x \, \varphi \left(\frac{x}{x_0} + \alpha_1\right)  \nonumber  \\
& & =A_1 A_2 x_0^2 \int_{\alpha_1}^\infty (z - \alpha_1) \varphi(z) \varphi(z-\alpha_1+\alpha_2) . \nonumber
\end{eqnarray}
To compute the correction due to the momentum dependent term one needs
\begin{eqnarray}
\langle2|p^2|1\rangle & = & - \hbar^2 \int_0^\infty dx \psi_2 (x) \frac{d^2}{dx^2} \psi_1 (x)  \label{symmp} \\
& = & - \hbar^2 A_1 A_2 \int_{0}^\infty d x \, \varphi \left(\frac{x}{x_0} + \alpha_2\right) \frac{d^2}{dx^2} \, \varphi \left(\frac{x}{x_0} + \alpha_1\right)  \nonumber \\
& = & + \hbar^2 A_1 A_2  \int_{0}^\infty d x \, \frac{d}{dx} \varphi \left(\frac{x}{x_0} + \alpha_2\right) \frac{d}{dx} \, \varphi \left(\frac{x}{x_0} + \alpha_1\right)  \nonumber \\
& = &  \frac{A_1 A_2}{x_0} \int_{\alpha_2}^\infty z  \varphi(z) \varphi(z-\alpha_2+\alpha_1). \nonumber
\end{eqnarray}
where the third line comes from partial integration, and we used $\varphi(+\infty) = 0 = \varphi (\alpha_2)$.

By combining the previous results one gets
\begin{eqnarray}
\hbar \Omega &=& (m_1 g_0 +m_0 g_1)x_0 I_1 \nonumber \\
&+& \frac{\hbar^2}{2 m_0} m_1 (1 / x_0^2) I_2,
\end{eqnarray}
with $I_1,I_2$ in Eqs.(\ref{I2}), which, after simple algebra, gives Eq.(\ref{Omega}).

Notice the explicit symmetry $ 1 \leftrightarrow 2$ of the second line of the expression (\ref{symmx}) and of the third line of the expression (\ref{symmp}), that proves $P_{12} = P_{21}$.

\end{document}